\documentclass[aps,showpacs,floatfix,twocolumn,superscriptaddress]{revtex4}
\usepackage{graphicx} 
\usepackage{dcolumn} 
\usepackage{amsmath} 

\newcommand{\be}{\begin{equation}}
\newcommand{\ee}{\end{equation}}
\newcommand{\bea}{\begin{eqnarray}}
\newcommand{\eea}{\end{eqnarray}}

\def\slr#1{\setbox0=\hbox{$#1$}                 
   \dimen0=\wd0                                 
   \setbox1=\hbox{/} \dimen1=\wd1               
   \ifdim\dimen0>\dimen1                        
      \rlap{\hbox to \dimen0{\hfil/\hfil}}      
      #1                                        
   \else                                        
      \rlap{\hbox to \dimen1{\hfil$#1$\hfil}}   
      /                                         
   \fi}

\begin{document}
\title{Infrared and Ultraviolet QCD  dynamics with quark mass for J=0,1 mesons. }

\author{N.~Souchlas}
\affiliation{Department of Physics,  Brookhaven National Lab, Upton NY 11973}
\email{nsouchlas@bnl.gov}

\begin{abstract}

By using a previously developed  phenomenological kernel for the study of the light quark QCD sector and dynamical 
  chiral  symmetry breaking effects  we will examine the relative infrared and ultraviolet QCD dynamics for J=0,1 meson 
properties. For the same reasons we  extend and explore   a quark mass depended generalization of the kernel in the  heavy quark 
region and we also compare with the original  model. The  relation between the dynamics of the quark propagator and the
 effective  kernel  with the J=0,1 QQ and qQ mesons and quarks Compton size is also discussed.

\end{abstract}



%


\preprint{Kent State U preprint no. KSUCNR-204-03}
\maketitle

\section{\label{sec:intro} Introduction}
A lot of effort has been focused on studying the spectrum and the properties of 
light quark mesons (see \cite{GutierrezGuerrero:2010md}-\cite{Maris:2006ea}) and 
references therein). In these systems non-perturbative 
effects are dominant, therefore they are the best candidates for understanding 
the mechanisms underlying confinement and dynamical chiral symmetry breaking which are fundamental 
elements and of crucial importance for the theory.
In some of  these studies the quark propagator equation  has provide useful insights in the light quark sector of  
QCD \cite{Bhagwat:2006xi}, \cite{Chang:2006bm} . 
It is also interesting to explore, using that fundamental block of the theory and the quarks bound state equation,
the transition to heavy quark physics. 

We plan  to canvass, by using an effective kernel for the gap equation, how the quark mass  affects 
the  infrared and ultraviolet dressing  of the propagator and how that in turn  will alter the dynamics 
of the bound state of quarks. A more realistic case of  a quark mass dependent version of the effective kernel is
also  explored in the same light. The Compton size of quarks and mesons is also used to 
qualitatively understand the relative infrared (IR) and ultraviolet (UV) QCD dynamics and
inspired an approach that enabled us to reach a physical bound state for $qb$ q=u/d,s,c mesons.
Aspects related to the finite size of hadrons involving  
recent ideas by   Brodsky and Shrock   \cite{Brodsky:2008be}  are also briefly discussed.

\section{Quark propagator, mesons bound state equations and Rainbow-Ladder truncation. }

The  Dyson-Schwinger equation for the quark propagator (gap equation) 
has the form: 

\begin{eqnarray}
 S(p)^{-1}= \; Z_{2} (i \not\!p\,+ \; m_{bm} )  \nonumber \\ 
+Z_{1}  \int_{q}^{\Lambda}   g^{2}D_{\mu \nu}(k) \gamma_{\mu} \frac{\lambda^{i}}{2} S(q) \Gamma_{\nu}^{i}(p,q)    
  \label{eq:dse}
\end{eqnarray}
\noindent  $D_{\mu \nu}(k)$ is the renormalized dressed gluon propagator, $\Gamma_{\nu}^{i}(p,q)$ is the renormalized
 dressed quark-gluon vertex, $\Lambda$ is the regularization mass scale, 
with $Z_{1} ,Z_{2}$ being  the gluon and quark propagator renormalization constants.
 
  Using the  rainbow truncation for the gap  equation and introducing  $\alpha(q^2)$:
\begin{eqnarray}
\lefteqn{
Z_1 g^2 D_{\mu \nu}(q) \Gamma^i_\nu(p,q) \to }
\nonumber \\ &&
 4\pi\,\alpha(q^2) \, D_{\mu\nu}^{\rm free}(q)\, \gamma_\nu
                                        \textstyle\frac{\lambda^i}{2} \,.
\label{Eq:rainbow}
\end{eqnarray} 
where $D_{\mu\nu}^{\rm free}(q)$ is the  free gluon propagator,
we can disentangle the equation from the rest of the Dyson-Schwinger equations.
 
The unrenormalized quark self-energy term of the gap equation in the rainbow truncation is: 
\begin{eqnarray}
~~~~~ \Sigma^{'}(p)= i \not\!p\,\{ A^{'}(p^{2})-1\} + B^{'}(p^{2}) = \nonumber  \\
 \frac{4}{3} \int^{\Lambda} \frac{d^{4}q}{(2\pi)^{4}}\frac{ \mathcal{G}(k^{2})}{k^{2}} T_{\mu \nu}(k) \gamma_{\mu} S(q)
 \gamma_{\nu},   \label{quarkselfenergy}
\end{eqnarray}
where we have set $ \mathcal{G}(k^{2}) =4 \pi \alpha(k^{2})$, $k=p-q$ is the gluon momentum  and the factor $ \frac{4}{3}$
 comes from the trace over the color indexes.
By taking the Dirac trace of the last equation we get:
\begin{eqnarray}
~~~   B^{'}(p^{2}) = 4 \int^{\Lambda} \frac{d^{4}q}{(2\pi)^{4}}\frac{\mathcal{G}(k^{2})}{k^{2}} \sigma_{s}(q^{2}),  
\label{Gapeq2}
\end{eqnarray}
and if we  multiply by  $\not\!p $ and then take the Dirac trace,  we get the second equation:
\begin{eqnarray}
~~~ p^{2} (A^{'}(p^{2})-1) =\frac{4}{3}  \int^{\Lambda} \frac{d^{4}q}{(2\pi)^{4}}\frac{\mathcal{G}(k^{2})}{k^{2}}\times
 \nonumber  \\
\bigg(p.q+2 \frac{(k.p)(k.q)}{k^{2}} \bigg)\sigma_{v}(q^{2}) , \label{Gapeq1}
\end{eqnarray}
where we have introduced the quark propagator amplitudes $\sigma_{s}(q^{2})$, $\sigma_{v}(q^{2})$:
\begin{eqnarray}
\sigma_{s}(q^{2})=\frac{1}{A(q^{2})}~ \frac{M(q^{2})}{q^{2}+M^{2}(q^{2})}~~~(a),\nonumber  \\ \sigma_{v}(q^{2})=
\frac{1}{A(q^{2})}~ \frac{1}{q^{2}+M^{2}(q^{2})}~~~(b).   \label{qpramplitudes}
\end{eqnarray}
The quark propagator in terms of $A^{'}$, $B^{'} $ is then: 
\begin{eqnarray}
S^{-1}(p)=i \not\!p\, A(p^{2})+ B(p^{2})= \nonumber  \\  Z_{2}(i \not\!p+ m_{bm})+\Sigma^{'}(p)= \nonumber  \\
i \not\!p(Z_{2}+A^{'}(p^{2})-1)+(m_{bm}+B^{'}(p^{2}))~~~~~   \label{qprAB}
\end{eqnarray}
Using  the propagator  renormalization condition,  $S^{-1}(p)|_{p^{2}=\mu^{2}}= i \not\!\mu \, +m_{r}(\mu^{2})$, we get 
\begin{eqnarray}
A(\mu^{2},\Lambda^{2})=1+A^{'}(p^{2},\Lambda^{2})-A^{'}(\mu^{2},\Lambda^{2} ), \label{renormcond3}
\end{eqnarray}
\begin{eqnarray}
B(\mu^{2},\Lambda^{2})=m_{r}(\mu^{2}) +B^{'}(p^{2},\Lambda^{2})-B^{'}(\mu^{2},\Lambda^{2} ). \label{renormcond4}
\end{eqnarray}
where $ m_{r}(\mu^{2})$ is the renormalized current quark mass at point $\mu=19~GeV$ and  it is a  parameter we 
 fit to experimental data.

 The amplitude (BSA)  $\Gamma^{ab}_{M}(p,P)$  for a meson  state of  quarks of flavors a and b is given from the 
solution of   the  homogeneous Bethe-Salpeter equation (BSE): 

\begin{eqnarray}
[\Gamma^{ab}(p,P)]_{tu}= \int^{\Lambda}\frac {d^{4}\tilde{q}}{(4\pi)^{4}} K_{tu}^{rs}(p,\tilde{q},P) \nonumber\\ 
\times [S^{a}(\tilde{q}+\eta P)
\Gamma^{ab}(\tilde{q},P) S^{b}(\tilde{q}-\bar{\eta}P)]_{sr}    \label{eq:bse}
\end{eqnarray}
\noindent P is the total momentum,  $\eta$ ($\bar{\eta}$) is the momentum partitioning parameter for the quark (antiquark)
 and    $\eta+\bar{\eta}=1$, $\eta \in[0,1]$. 
$K_{tu}^{rs}(p,\tilde{q},P)$ is the unknown renormalized amputated irreducible  quark-antiquark scattering kernel. Physical
 observables are independent of the partitioning parameter.

   The most general form of the BSA for psudoscalar mesons has four invariants
while for the vector mesons has eight (see \cite{Maris:1999nt},\cite{Maris:2003vk}) and we use a four Chebychev polynomial
  expansion  for each one of them.
These amplitudes  are Lorentz  scalar functions of $q^{2}$, $P^{2}$, $q.P$ and the  momentum partitioning parameter $\eta$.
 For qQ mesons that parameter help us avoid   having the   propagator singularities inside their mass shell BSE integration domain.
Since for the   mass shell  momentum $ P^{2}=-m^{2}$, where  $ m$  is the meson mass,  the quark momenta in
 (eq. ~\ref{eq:bse}) are in general complex numbers. This requires the solution  of  the gap equation in the 
appropriate parabolic region in the complex plane.

 The ladder truncation for the BSE is an approximation for the equation's kernel:
\begin{eqnarray}
\lefteqn{ [K(p,q,P)]^{rs}_{tu} \to }
\nonumber \\ &&
        -4\pi\,\alpha(q^2)\, D_{\mu\nu}^{\rm free}(q)
        \textstyle[{\frac{\lambda^i}{2}}\gamma_\mu]^{ru} \otimes
        \textstyle[{\frac{\lambda^i}{2}}\gamma_\nu]^{ts} \,,
\label{Eq:ladder}
\end{eqnarray}

The  electroweak decay constant $f_{H}$ of a charged pseudoscalar meson  \cite{Maris:1999nt}
 expressed in terms of the meson normalized BSA and quark propagators:
\begin{eqnarray}
f^{PS}_{H}=\frac{Z_{2}N_{C}}{P^{2}}\times  \nonumber \\ 
 \bigg\{\int^{\Lambda} \frac{d^{4}q}{(2\pi)^{4}} P_{\mu} {\rm Tr}_{D} \big[\Gamma^{ab}_{M}(q,P) S^{b}(q_{-}) \gamma_{\mu} 
\gamma_{5} S^{d}(q_{+})\big]\bigg\}
\label{eq:decayconstps} 
\end{eqnarray}
where $P^{2}=-m^{2}_{H}$ and $N_{C}=3$ is the number of colors, from  the trace over the color indexes.
Similar expression exists  for vector mesons.

\section{Rainbow-Ladder effective interaction. }

For the unknown effective running coupling we are going to use a  kernel-model  that has been developed within the
 Rainbow-Ladder  truncation of Dyson-Swchwinger 
equations. The model respects some of the most important symmetries of QCD, like Chiral symmetry and Poincare covariance, 
 while it  provides quark dressing,
dynamical Chiral symmetry breaking, and most important,  quark confinement.   
It has been   used  to study the physics of DCSB and related phenomena, like the spectrum  of light quark mesons 
(\cite{Maris:1999nt} \cite{Maris:2003vk}, \cite{Maris:1997tm}), decay constants (\cite{Maris:2003vk}, 
\cite{Maris:1997hd}, \cite{Holl:2004fr}, \cite{Jarecke:2002xd}, \cite{Krassnigg:2004if}) and other physical observables 
 (\cite{Roberts:1994hh}, \cite{Maris:2000sk}, \cite{Maris:1999bh}),  in  good agreement with experimental data
(\cite{Holl:2005vu}, \cite{Maris:1998hc}, \cite{Maris:2000wz}, \cite{Maris:2002mz}).

The so called Maris-Tandy (MT) model \cite{Maris:1999nt}  has the form:  
\begin{eqnarray}
\frac{4 \pi \alpha(k^{2})}{k^{2}} = \frac{(2\pi)^{2} k^{2}D} {\omega^{6}} e^{-\frac{k^{2}}{\omega^{2}}}  
 +\frac{2 (2\pi)^{2}  \gamma_{m} F(k^{2})}{\ln[\tau+(1+\frac{k^{2}}{\Lambda^{2}_{QCD}})^{2}]}  
\label{Eq:MTmodel}
\end{eqnarray}
For the parameters we  have    $\omega=0.4~GeV ,D=0.93~GeV^{2}$ and $m_{t}=0.5~GeV$ and the u/d- and s-current 
quark masses at the renormalization scale $\mu=19~GeV$, fitted to the experimental masses of pion and kaon, are
 $m_{u/d}=0.00374~GeV$ and 
$m_{s}=0.083~GeV$ . For the other two quarks, we use the masses  in \cite{Maris:2005tt}, where 
$m_{c}=0.88~GeV$ and $m_{b}=3.8~GeV$. 
The model essentially simulates both gluon and quark-gluon vertex dressing effects. 
The phenomenological first term defines the behavior  in the infrared region
and provides the infrared enhancement necessary  for the right value of the quark condensate in the chiral limit. 
The second term   is important 
in the ultraviolet region and is set up so that will  reproduce the 1-loop perturbative QCD running coupling  behavior.

\section{Heavy  quark propagator within meson dynamics.}
Recently  an effort has started
 to extend the  applications of the MT model  in the heavy quarks region (\cite{Maris:2006ea},  \cite{Krassnigg:2009zh}   \cite{nick:2009}).
Due to its behavior  in the complex domain the solution of the heavy quark gap equation is challenging.
By changing the integration variable in the  equation, from the quark internal momentum to the gluon momentum, 
solved that problem and the only limitation comes from the propagator's singularities. The type and  the exact 
location of these points is 
not known. 
Notice that the thranslationally invariant regularization of the integrals allows that change in the integration variable and variations 
of parameter $\eta $ and that our approach to numerically solve the equation is different than that in \cite{Krassnigg:2009zh}.

\begin{figure}
\centering
\includegraphics[height=5cm,width=8cm]{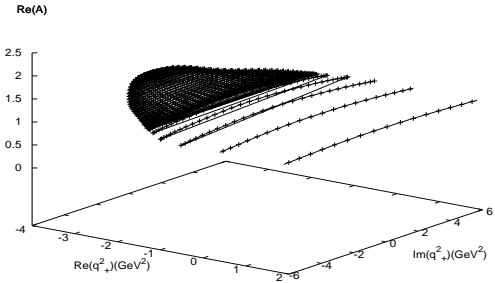}
\caption{$Re(A)$ for c quark in a parabolic region in the complex plane ($ P^{2}=-14~GeV^{2}$, $\eta$=0.50). In 
this plot and in the following ones the straight lines connecting the end-points have no significance and should be ignored.}
\label{fig:Real_A_kcp_32_32_32_32_E-7_P2-14_n_0.5_0.88}
\end{figure}

\begin{figure}
\centering
\includegraphics[height=5cm,width=8cm]{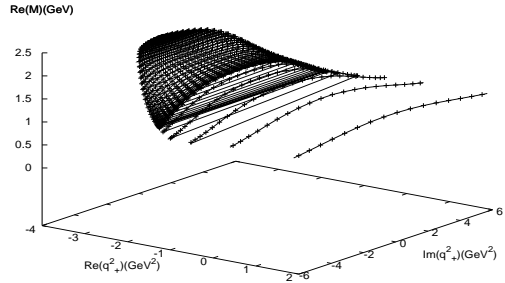}
\caption{$Re(M)$ for  c quark in a parabolic region in the complex plane ($ P^{2}=-14~GeV^{2}$, $\eta$=0.50).}
\label{fig:Real_M_kcp_32_32_32_32_E-7_P2-14_n_0.5_0.88}
\end{figure}

\begin{figure}
\centering
\includegraphics[height=5cm,width=8cm]{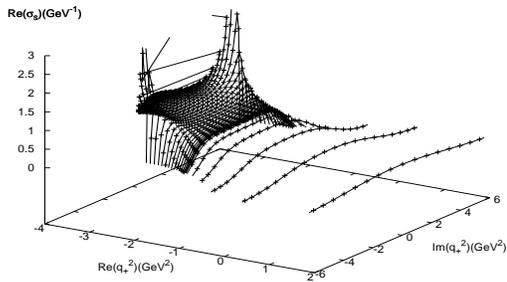}
\caption{$Re(\sigma_{s})$ for  c quark in a parabolic region in the complex plane ($ P^{2}=-14~GeV^{2}$, $\eta$=0.50)
 with integration over the gluon momentum (kcp).  The peak of the parabolic region  is at $ (\eta P)^{2}=-3.5~GeV^{2}$.  
We see  indications  of the existence of a pair of complex conjugate singularities near the peak of the region, 
approximately located at ($x_{o}$, $y_{o}$)$\sim$(-2.7, $\pm$3.5) $GeV^{2}$. The type of the singularities is unknown.}
\label{fig:Real_sigma_s_kcp_32_32_32_32_E-7_P2-14_n_0.5_0.88}
\end{figure}

In the case of the c quark for example with  current mass $m_{c}=0.88 ~GeV$ and for the parabolic region determined by 
 $q^{2}_{+}=(\tilde{q}+\eta P)^{2}=\tilde{q}^{2}-(\eta M)^{2}+2i\eta\sqrt{\tilde{q}^{2}M^{2}} v  $ where $\tilde{q}$ is the BSE 
integration variable, with $\eta=0.50$ and $ P^{2}=-M^{2}=-14~GeV^{2}$ (the peak of the region will be 
then at $(\eta P)^{2}=-3.5~GeV^{2}$) amplitudes  $Re(A)$, $Re(M)$ and $Re(\sigma_{s})$ are  plotted in
 Figures (\ref{fig:Real_A_kcp_32_32_32_32_E-7_P2-14_n_0.5_0.88},~\ref{fig:Real_M_kcp_32_32_32_32_E-7_P2-14_n_0.5_0.88},
~\ref{fig:Real_sigma_s_kcp_32_32_32_32_E-7_P2-14_n_0.5_0.88}) correspondingly. For our present studies of c-quark mesons 
we need to solve the gap equation for $ P^{2}$ as small as $ P^{2}=-M_{V}^{2}\sim -9.6~GeV^{2}$, so the mass 
shell point (peak) is at $(\eta P)^{2}\sim -2.4~GeV^{2}$ and the singularities are far  from the BSE integration
 domain and they present no problem. This is also  tested by varying parameter  $ \eta $.


 From these plots we can see that only amplitude $\sigma_{s}$  (and the same is true for  $\sigma_{v}$) has singularities, 
but not A and M.  We can conclude then that the singularity is the point where the denominator of $\sigma_{s/v}$ vanishes,
 i.e.  $q^{2}_{+}+M^{2}(q^{2}_{+})=0$.

\begin{figure}
\centering
\includegraphics[height=5cm,width=8cm]{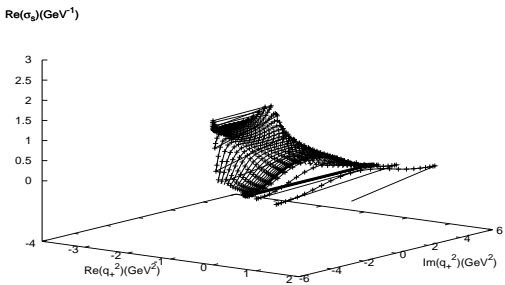}
\caption{$Re(\sigma_{s})$ for  c quark in a parabolic region in the complex plane ($ P^{2}=-8~GeV^{2}$, $\eta$=0.50)  
keeping only the first IR term of the kernel. The peak of the parabolic region is  at $ (\eta P)^{2}=-2.0~GeV^{2}$. The 
solution is different but we still have the general characteristics of the solution from the  complete MT model. The first
 pair of singularities  approximately appear to be ($x_{o}$, $y_{o}$)$\sim$(-1.7,$\pm$2) $GeV^{2}$. }
\label{fig:Re_sigma_s_1st_MT_term_kcp_32_32_32_32_E-7_P2-8.0_n_0.5_0.88}
\end{figure}

\begin{figure}
\centering
\includegraphics[height=5cm,width=8cm]{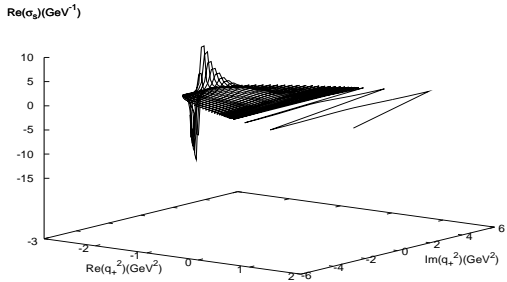}
\caption{$Re(\sigma_{s})$ for the c quark in a parabolic region in the complex plane ($ P^{2}=-8~GeV^{2}$, $\eta$=0.50) 
with integration over the gluon momentum (kcp) and keeping only the UV  term in the model. The peak of the parabolic 
region is at $ (\eta P)^{2}=-2.0~GeV^{2}$. The first singularity is on the real axis around 
($x_{o}$, $y_{o}$)$\sim$(-1.9, 0) $GeV^{2}$. }
\label{fig:Re_sigma_s_2st_MT_term_kcp_32_32_32_32_E-7_P2-8.0_n_0.5_0.88}
\end{figure}

By keeping only the infrared  first term of the model the solution  for $Re(\sigma_s)$ for the same quark is
 plotted in fig. (\ref{fig:Re_sigma_s_1st_MT_term_kcp_32_32_32_32_E-7_P2-8.0_n_0.5_0.88}). 
In the case where we keep only the UV  term in the MT model, the solution has the first  singularity on  the real axis 
(see fig. \ref{fig:Re_sigma_s_2st_MT_term_kcp_32_32_32_32_E-7_P2-8.0_n_0.5_0.88}). 

Comparing  the c and  b quark real part of the mass amplitude Re(M), the last one  appears to be almost flat near 
the peak of the parabolic region.
Re(A) also appears to vary little, just above one, in the complex plane. One may consider that such behavior of Re(M) and Re(A) justifies 
for a constituent like approximation for the b quark propagator.    From the b quark  Re($\sigma_{s}$) plot we found 
 the singularities have not only moved, as was expected, deeper in the time-like region, along the real  axis, but also 
further apart along the imaginary axis. This is  exactly the opposite from a constituent-like  behavior of a propagator. 
That also signifies the importance of the imaginary part of the dressed quark  mass amplitude in the behavior of the 
propagator amplitudes $\sigma_{s}$ and $\sigma_{v}$.

\begin{figure}
\centering
\includegraphics[height=5cm,width=8cm]{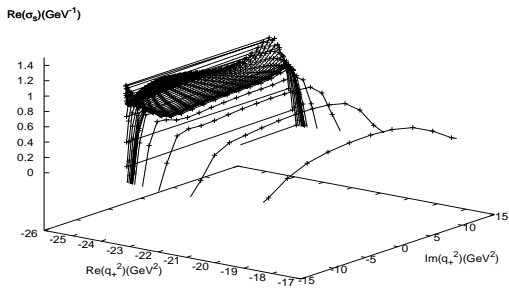}
\caption{$Re(\sigma_{s})$ for the b quark in a parabolic region in the complex plane ($ P^{2}=-101~GeV^{2}$, $\eta$=0.50)
 with integration over the gluon momentum (kcp) with the full MT model . The peak of the parabolic region is at 
$ (\eta P)^{2}=-25.25~GeV^{2}$. Traces  of a first pair of singularities can be seen near the peak of the region and
 they are approximately located at ($x_{o}$, $y_{o}$)$\sim$(-24.0,$\pm$12) $GeV^{2}$. }
\label{fig:Real_sigma_s_full_MT_kcp_32_24_24_16_E-6_P2-101_n_0.5_mb_3.8}
\end{figure}

In Table \ref{Table:Full_IR_UV_approximate_singularities_location} we collect the data for the approximate location ($x_{o}$,$y_{o}$) of the
 singularities for the c and b propagators for the full, IR and UV gap  solution. In that table we include a qualitative 
estimation of the quark mass dressing for each case. We assume that the location of the singularities along the real axis
 can be used to extract that piece of information. From these data we observe there is an increase in the quark 
mass dressing as we go from  c to  b quark using the full model. For  the c quark there is an almost equal
 dressing from the IR and UV term of the interaction while for   b quark the UV term provides more than twice the 
mass dressing of the IR term. 
\begin{table*}
\caption{c and b quark propagator approximate location of the singularities in the complex plane for the full MT, IR and
 UV model gap solution. All data are in GeV units. For the qualitative  approximate  estimation of the quark mass effective
  dressing we assume that $(m_{Q}(19~GeV)+ M^{Q}_{\Sigma})^{2}\sim -x_{o}$.
\label{Table:Full_IR_UV_approximate_singularities_location} }
\begin{center}
\begin{tabular}{|l||l||l||l|l||l|l|} \hline \hline
           & \multicolumn{2}{|c||}{full MT} & \multicolumn{2}{|c||}{IR term} & \multicolumn{2}{|c|}{UV term} \\ \hline \hline 
  quark    & $(x_{o},y_{o})$  & $M_{\Sigma}$ & $(x_{o},y_{o})$   & $M_{\Sigma}$  & $(x_{o},y_{o})$ & $M_{\Sigma}$         
   \\  \hline  \hline     
  $c$      &(-2.7,$\pm$3.5) & 0.763        & (-1.7,$\pm$2.0) & 0.424         &(-1.9,0.0)     &  0.498     \\  \hline \hline  

  $b$      &(-24.0,$\pm$12.0) & 1.1        & (-17.2,$\pm$7.5) & 0.347        &(-23.0,0.0)    &  0.996     \\  \hline \hline   
\end{tabular}
\end{center} 
\end{table*}

\section{Results for quarkonia  mesons.}
The results for the masses and the decay constants of the three pseudoscalar and vector quarkonia ($Q\bar{Q}$ Q=s,c,b)  
mesons  are collected in Tables (\ref{Table:ss1_PS_kcp_new_meth_meson_masses_decays_D_w}) and
 (\ref{Table:ss1_V_kcp_new_meth_meson_masses_decays_D_w}).
\begin{table}
\caption{$s\bar{s}$(fictional), $c\bar{c}$ and $b\bar{b}$   pseudoscalar  meson masses and decay constants  with their
 relative percentage differences from calculations where only the first  infrared (IR) term or only the ultraviolet(UV) 
perturbative tail in the MT model is retained.
\label{Table:ss1_PS_kcp_new_meth_meson_masses_decays_D_w} }
\begin{center}
\begin{tabular}{|l||l||l|l||l|l|} \hline \hline
  \multicolumn{6}{|c|}{$s\bar{s}$,$c\bar{c}$ and $b\bar{b}$ pseudoscalar meson masses}  \\  \hline \hline 
  meson     & full MT & IR only   & $\Delta M/M\%$ &UV only&$\Delta M/M\%$              \\  \hline  \hline     
  $s\bar{s}$& 0.696   & 0.565     & $-18.7$        &--             &--                  \\  \hline \hline   
 
  $c\bar{c}$& 3.035   & 2.41      & $-20.6$        &--             &--                  \\  \hline \hline   
   
  $b\bar{b}$& 9.585   & 8.106     & $-15.4$        & 9.530         & $-0.6$             \\  \hline \hline   \hline

  \multicolumn{6}{|c|}{$s\bar{s}$,$c\bar{c}$ and $b\bar{b}$ pseudoscalar meson decay constants}      \\  \hline \hline 
  meson     & full MT & IR only   & $\Delta f/f\%$ &UV only&$\Delta f/f\%$                           \\  \hline  \hline     
  $s\bar{s}$& 0.182   & 0.151     & $-17.0$       &--             &--                         \\  \hline \hline   
 
  $c\bar{c}$& 0.387   & 0.276     & $-28.7$       &--             &--                         \\  \hline \hline   
   
  $b\bar{b}$& 0.692   & 0.172     & $-75.1$       & 0.606         & $-12.4$                    \\  \hline \hline   

\end{tabular}
\end{center}
\end{table}
\begin{table}
\caption{$s\bar{s}$, $c\bar{c}$ and $b\bar{b}$   vector   meson masses and decay constants  with their relative percentage differences from calculations where only the first  infrared (IR) term or only the ultraviolet(UV) perturbative tail in the MT model is retained.
\label{Table:ss1_V_kcp_new_meth_meson_masses_decays_D_w} }
\begin{center}
\begin{tabular}{|l||l||l|l||l|l|} \hline \hline
  \multicolumn{6}{|c|}{$s\bar{s}$,$c\bar{c}$ and $b\bar{b}$ vector meson masses}        \\  \hline \hline 
  meson     & full MT & IR only   & $\Delta M/M\%$ &UV only&$\Delta M/M\%$              \\  \hline  \hline     
  $s\bar{s}$& 1.072   & 0.949     & $-11.5$        &--             &--                  \\  \hline \hline   
 
  $c\bar{c}$& 3.235   & 2.588     & $-20.0$        &--             &--                  \\  \hline \hline   
    
  $b\bar{b}$& 9.658   & 8.130     & $-15.9$        & 9.586         & $-0.8$             \\  \hline \hline   \hline

  \multicolumn{6}{|c|}{$s\bar{s}$,$c\bar{c}$ and $b\bar{b}$ vector meson decay constants}      \\  \hline \hline 
  meson     & full MT & IR only   & $\Delta f/f\%$ &UV only&$\Delta f/f\%$                     \\  \hline  \hline     
  $s\bar{s}$& 0.259   & 0.274     & $+5.8$        &--             &--                          \\  \hline \hline   
 
  $c\bar{c}$& 0.415   & 0.333     & $-19.8$       &--             &--                          \\  \hline \hline   
   
  $b\bar{b}$& 0.682   & 0.340     & $-50.1$       & 0.510         & $-25.2 $                   \\  \hline \hline   

\end{tabular}
\end{center}
\end{table}
For the fictional pseudoscalar $s\bar{s}$, the IR term  gave a mass that is smaller by 18.7 \%
from that of the full model , while for the decay constant we have  a 17.0 \% decrease. 
For the  $c\bar{c}$ meson it appears the  first infrared term accounts for about 80 \% of the meson mass
and about 70\% for the decay constant. Additional studies show an equal sensitivity of the decay constant 
for the $s\bar{s}$ on quark's  UV dressing effects, directly and  indirectly through the BSA,
while for the $c\bar{c}$ it  appears to be  more sensitive to UV corrections in the meson
 amplitude.  No bound state could be reached for these two systems with  the UV tail term only.

 For the b quarkonium we  first observe  that now  even the very weak, tail term, of the interaction can give us a bound state. 
Therefore a very weak attractive force is adequate for such heavy particles to bind them together. 
The mass of the system is lighter by just  0.055 GeV or   0.57 \% of the mass from the full model, while the decay constant, 
more sensitive to dressing details,  decreases by 12.4 \%, when we keep  the UV tail term only. 
On the other hand, since the IR term is much more attractive  than the  long distance  term, we get a much smaller mass,
a 15.4\% relative decrease,  while the decay constant decrease by more than 75 \%.
This also signifies a dominance of the UV tail term over the IR long distance term, supported by the heavy quark 
mass behavior of the propagator. 

Same observations are also true for the corresponding vector mesons data in table  
\ref{Table:ss1_V_kcp_new_meth_meson_masses_decays_D_w}. The only difference we notice is the greater gap  between the
 full model calculated decay constant for the  b quarkonium and the one calculated by using only  the UV term . The percentage
 difference is about twice as much that of the corresponding pseudoscalar meson and the reason is that  vector mesons
 are more extended objects than  pseudoscalar mesons, hence  the IR term contributes more in the evaluation of the
 physical observables. The decay constant indicates a deficiency of the low momenta behavior of the UV  term
 on the quarks relative angular momentum.

\begin{figure}
\centering
\includegraphics[height=5cm,width=8cm]{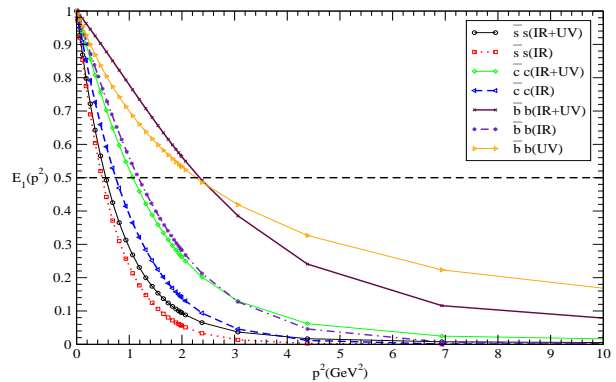}
\caption{First Chebychev moment of the dominant  invariant for the $s\bar{s}$, $c\bar{c}$, $b\bar{b}$  pseudoscalar mesons,
 from the full MT model solution and the solution with only the IR term. For the  $b\bar{b}$  quark system we also have an 
UV calculated amplitude since we can  reach a mass shell even  with only the weaker UV tail term in the interaction.}  
\label{fig:E1_kcp_MT_IR_UV_ss_cc_bb_24-24-24-24-16_n_0.50}
\end{figure}

The first Chebychev moment of the dominant  invariant for the equal quark pseudoscalar mesons  from the different 
approaches  appear  in fig.  (\ref{fig:E1_kcp_MT_IR_UV_ss_cc_bb_24-24-24-24-16_n_0.50}). 
Similar behavior is observed for the corresponding vector mesons.

As the mass increases we observe that the IR amplitude has a   faster  decrease with momentum 
than the full model amplitude and the system is more delocalised, consistent with the expectations of a relative smaller 
 dressing of the quark mass provided by the IR term.   The  UV amplitude of the b  quarkonium on the other hand 
decreases  slower than the  MT one and  the system is slightly more localized. This is a sign that for this 
meson the UV short distance term is  now providing most of the  quark mass dressing and binding force for the two quarks.

\section{Relative infrared and ultraviolet QCD dynamics for quarkonia.}

\begin{figure}
\centering
\includegraphics[height=5cm,width=8cm]{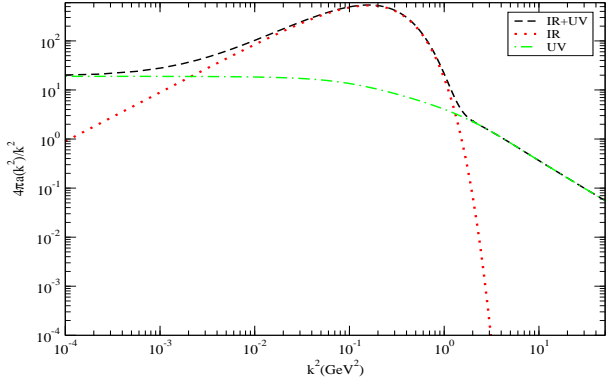}
\caption{Plot of the full MT model and the model with only the IR or UV tail term. The IR term will mostly determine 
 the strong behavior of the model for $k^{2}<2$ $GeV^{2}$ and beyond that point the UV term dominates.}  
\label{fig:MT_FULL_IR_UV_function_new_real_axis_D_w}
\end{figure}

From the last results become obvious that as the current quark mass increases the IR term has a marginalized   important in the
 realization of  meson observables. The question is why and how that happens.  The effective kernel  is the same and there is no 
change in the UV contribution in the quarks interaction. So essentially this is a result of the effect of the large 
current quark mass in the behavior of the propagator amplitudes and consequently the combined dynamical interplay with the
 MT effective kernel. A first naive qualitative explanation is that as we raise the quark mass the amplitudes will have smaller values,
 even very deep in the time-like region, and we need to get  closer and closer  to their  singularities to notice some important 
increase in their values.
That in turn will suppress the contribution and lessen the significance of the IR term while the UV term of the model
 will become more relevant in the evaluation of the observables. That  is also  related  to the extremely fast decay of 
the IR term and much slower decrease of the UV one (see fig. \ref{fig:MT_FULL_IR_UV_function_new_real_axis_D_w}).

\subsection{A first look in the quark mass dynamics of the propagator with the MT kernel.}

\begin{figure}
\centering
\includegraphics[height=5cm,width=8cm]{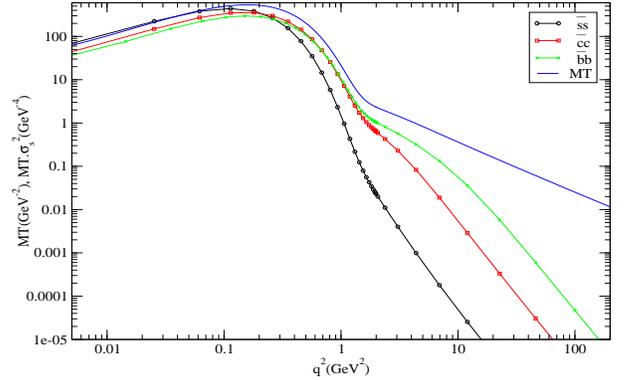}
\caption{Plot of the product  MT$\cdot\sigma^{2}_{s}$ for the $s\bar{s}$, $c\bar{c}$, $b\bar{b}$ quark systems compared 
with the MT model behavior  for external  relative quark momentum $p^{2}=0$. In all cases the propagator amplitude along the real axis
 is for the on-shell solution of the BSE.  One can clearly see the \textit{increasing} suppression imposed by the 
amplitude in the IR region and the \textit{decreasing} suppression on the UV tail term,  as the current quark mass
 \textit{increase}. This IR suppression is mild compared to the other two cases,
 due to the presence of  the dressed mass  in the numerator of $\sigma_{s}$, moderating the decrease of the amplitude. 
Same things are true for every  $p^{2}>0$ but there is a significant scale down as $p^{2}$ increase.}  
\label{fig:MT.sigma_s_2_new_ss_cc_bb_ms_0.083_mc_0.88_mb_3.8}
\end{figure}

\begin{figure}
\centering
\includegraphics[height=5cm,width=8cm]{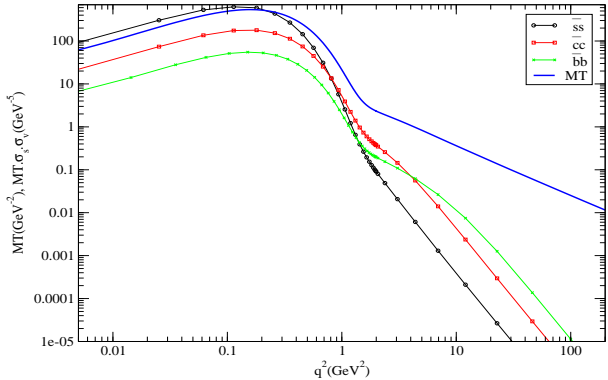}
\caption{Plot of the product  MT$\cdot\sigma_{s}\cdot\sigma_{v}$ for the $s\bar{s}$, $c\bar{c}$, $b\bar{b}$ quark systems 
compared with the MT model behavior  for external  momentum $p^{2}=0$. In all cases the propagator amplitude along the
 real axis is for the on-shell solution of the BSE.  As one would expect, from the two previous plots, the degree of 
the effects  in the IR and UV region is   somewhere  between the degree of the effects of the other two terms. Same 
things are true for every $p^{2}>0$ but there is a significant scale down as $p^{2}$ increase.}  
\label{fig:MT.sigma_s.sigma_v_new_ss_cc_bb_ms_0.083_mc_0.88_mb_3.8}
\end{figure}

To qualitatively see the effect of the heavy quark mass within the meson dynamics, we plot the product of the MT rainbow-ladder 
kernel  with the  two propagators amplitudes as they appear  in the BSE.   There are four  cases: 
$\sigma_{s}(q^{2}_{+})\cdot\sigma_{s}(q^{2}_{-})$, $\sigma_{v}(q^{2}_{+})\cdot\sigma_{v}(q^{2}_{-})$ and 
$\sigma_{s}(q^{2}_{\pm})\cdot\sigma_{v}(q^{2}_{\mp})$, but for convenience we consider only the real axis where we have
 Re($\sigma_{s/v}(Re(q^{2}_{\pm})))=\sigma_{s/v}(q^{2}-(\eta P)^{2})$, so there are essentially only three  amplitude 
products in the BSE integrand: $\sigma^{2}_{s}(q^{2}-(\eta P)^{2})$, $\sigma^{2}_{v}(q^{2}-(\eta P)^{2})$ and
 $\sigma_{s}(q^{2}-(\eta P)^{2})\cdot\sigma_{v}(q^{2}-(\eta P)^{2})$.  The plots of these products with the MT 
kernel for the $s\bar{s}$, $c\bar{c}$, $b\bar{b}$ pseudoscalar mesons  are in 
fig. (\ref{fig:MT.sigma_s_2_new_ss_cc_bb_ms_0.083_mc_0.88_mb_3.8}, 
 \ref{fig:MT.sigma_v_2_new_ss_cc_bb_ms_0.083_mc_0.88_mb_3.8}, 
\ref{fig:MT.sigma_s.sigma_v_new_ss_cc_bb_ms_0.083_mc_0.88_mb_3.8}).

\begin{figure}
\centering

\includegraphics[height=5cm,width=8cm]{MT.sigma_v_2_new_ss_cc_bb_ms_0.083_mc_0.88_mb_3.8.eps}
\caption{Plot of the product  MT$\cdot\sigma^{2}_{v}$ for the $s\bar{s}$, $c\bar{c}$, $b\bar{b}$ quark systems compared
 with the MT model behavior for external  momentum $p^{2}=0$ . In all cases the propagator amplitude along the real axis
 is for the on-shell solution of the BSE.  Once  again we  see the \textit{increasing} suppression imposed by the
 amplitude in the IR region and the same, for all cases after certain $q^{2}$,  suppression of the UV tail term,  as the 
current quark mass increase. $\sigma^{2}_{v}$ amplitude unlike $\sigma^{2}_{s}$ doesn't  have the quark mass amplitude  in the 
numerator, so we notice a much stronger suppression in the IR region  as the current quark mass
 increase. This suppression extents somehow in the UV region, but for $q^{2}>10~GeV^{2}$ is the same for all cases. 
Finally notice that for the s quark system we observe an \textit{enhancing} effect for the MT model in the IR region.
 Same things are true for every $p^{2}>0$ but there is a significant scale down as $p^{2}$ increase.}  
\label{fig:MT.sigma_v_2_new_ss_cc_bb_ms_0.083_mc_0.88_mb_3.8}

\end{figure}

We observe some mild \textit{increasing}  suppression in the IR region as the current quark mass \textit{increases}  
for the product  MT$\cdot\sigma^{2}_{s}$, and a much stronger \textit{increasing} suppression for the second  product 
due to the $\sigma^{2}_{v}$ amplitude. For the last product the situation is somewhere between the first two since it 
combines the effects of both propagator amplitudes. For the last two cases of products involving $\sigma_{v}$, we actually 
have a small  enhancement in the  IR region  for the s quark system, while from the MT$\cdot\sigma^{2}_{s}$    plot 
we notice $\sigma^{2}_{s}$ has almost no effect in the IR region. 
For the UV region in all cases we have suppression of the tail term. That suppression though
 is \textit{decreasing} as the current quark mass \textit{increases}, somehow faster for the MT$\cdot\sigma^{2}_{s}$
 than in the   MT$\cdot\sigma_{s}\cdot\sigma_{v}$ product. Finally for the term involving  $\sigma^{2}_{v}$ there is no
 difference for all quark systems for $q^{2}>10~GeV^{2}$ (the increasing IR suppression just extends a little beyond 
1 $GeV^{2}$ as the quark mass increase). The difference in the effects  between the first and the second type of terms 
are due to the fact that  $\sigma_{s}$ amplitude has also the quark mass amplitude in the numerator moderating the 
suppressing effects.

\subsection{Propagator amplitudes and quark dressing dynamics with quark mass.}

Although from these plots one can qualitatively and visually understand  why the IR term becomes  less important while the UV weak tail
 term becomes more relevant  for the mesons physical observables as we increase  the quark mass, the question is 
\textit{how}  that happens. For that reason we have to focus on the dynamics  of the two amplitudes in
 the two regions (IR and UV) and on how that  changes as we increase the current quark mass. The arguments will be  
again mostly qualitative.
From the two coupled gap equations   (\ref{Gapeq2}, \ref{Gapeq1}) and using the plots 
(\ref{fig:MT.sigma_s_2_new_ss_cc_bb_ms_0.083_mc_0.88_mb_3.8}, \ref{fig:MT.sigma_v_2_new_ss_cc_bb_ms_0.083_mc_0.88_mb_3.8}) 
we can extract some qualitative information about the relative change  the IR and UV tail term  will bring
 upon the  propagator amplitudes $M(p^{2})$ and $A(p^{2})$ as we increase  the  mass.  We notice that the integrand 
  of the equation for   $B^{'}(p^{2})$ is just  MT$\cdot\sigma_{s}$. That term  is almost unchanged in the IR region. 
There is only a small suppression over there as we increase the quark mass but  there is  a decreasing UV suppression. 
Therefore  it is \textit{possible} for the UV \textit{integrated  strength}  to dominate over 
the behavior of $B^{'}(p^{2})$. 
On the other hand MT$\cdot\sigma_{v}$  appears in the other integral equation. There is almost no change of that function 
in large momenta (larger than about 9-10 $GeV^{2}$) but up to this point there is very strong suppression
 greatly diminishing the contribution for the values of    $A^{'}(p^{2})$. As a result we expect  amplitude   $A(p^{2})$ 
to have smaller variation in the complex plane and get values closer to one as the quark mass increase. Amplitude  M (M=B/A)
 on the other hand, since it is proportional to B, will still be receiving important contributions from the 
self-interaction term and even for the heavy b quark will vary much and not very close to the b current mass of 3.8 GeV.
 Dressing IR effects will also  diminish and will be replaced by  UV dressing effects. 
So it should be $\Delta(M^{IR}_{\Sigma})<0,~\Delta(M^{UV}_{\Sigma})>0$.
 If one expects that  overall $\Delta(M_{\Sigma})<0$ then, since $M_{\Sigma}=M^{IR}_{\Sigma}+M^{UV}_{\Sigma}$, 
we should have $|\Delta(M^{IR}_{\Sigma})|>|\Delta(M^{UV}_{\Sigma})|$. Notice that the estimated quark mass 
dressing using the singularities location of  the IR and UV gap solutions for the c and b quarks 
(Table \ref{Table:Full_IR_UV_approximate_singularities_location}) do confirm the first two expected changes in the IR and
 UV  quark mass dressing contributions as the quark mass is raised, but don't confirm the third case, since from these
 data appears the UV dressing effects increase faster than the IR decrease and the  overall dressing increase. It is 
not clear though if this is because of a disadvantage of the MT model (IR term)  or because of our approach to estimate these dressings.
From the mass amplitude plots in section IV  and from the quarkonia results in section V  we have further evidence  that the 
short distance   dressing  \textit{increases}  in strength, \textit{faster} that the long  distance  dressing decreases,  
with the quark mass.  
Therefore it is  possible, after certain quark mass scale, that the  heavy quarks receive more  dressing from 
 the short distance than the long distance self-interaction. Also, at a different mass scale we may also have more sort-distance self dressing  than what   the lightest u/d quarks get from the long distance self interaction. 
Another element that supports the validity of the above ideas is  that the short   distance analysis is essentially model independent, 
dictated by the perturbative  QCD propagator and MT  tail term.   
As a final note, based on the above analysis, the heavy quark bound states and confinement of their quarks, are more a result 
of their large mass and their relatively larger UV dressing than a result of the binding interaction between them.  

\subsection{A qualitative mathematical analysis.}

Next we will try to qualitatively understand how a raising quark mass can have the effects we observed so far.
For the light quarks we assume $m_{q}<<\Lambda_{\chi}$ where $\Lambda_{\chi}\sim~1~GeV$ is the chiral symmetry breaking
 scale. . For the $\sigma_{s}$ amplitude we have  
\begin{eqnarray}
  \sigma_{s}(q^{2}-M_{H}^{2}/4)=  \nonumber \\  \nonumber \\ 
\frac{1}{A(q^{2}-M_{H}^{2}/4)}\frac{M(q^{2}-M_{H}^{2}/4)}{[q^{2}-M_{H}^{2}/4+M^{2}(q^{2}-M_{H}^{2}/4)]}= \nonumber \\  
\nonumber \\ 
\frac{1}{A(q^{2}-M_{H}^{2}/4)}\frac{m_{q}+M_{\Sigma}}{[q^{2}-M_{H}^{2}/4+m_{q}^{2}+M_{\Sigma}^{2}+2m_{q}M_{\Sigma}]}
\label{Eq:sigma_s_Real_q_approx}
\end{eqnarray}
 where $M_{H}$ is the hadron mass.  
First we will  focus on the IR region. We are very close to the peak of the parabolic region and for convenience of our 
analysis we take $q^{2}\sim 0$. For the light quarks we have strong dressing effects in that region  so
 $m_{q}+M_{\Sigma}\sim M_{\Sigma}$, and the propagator amplitude is approximately 
\begin{eqnarray}
\sigma_{s}(q^{2}\sim 0)\sim \frac{1}{A} \cdot \frac{M_{\Sigma}}{M_{\Sigma}^{2}-M_{H}^{2}/4}
\label{Eq:sigma_s_Real_q_approx_IR}
\end{eqnarray}
 Although A increases  near the peak, decreasing  the values of   $\sigma_{s}$, the hadron mass $M_{H}$
 is mostly due to the quark IR 
dressing effects and therefore  the  difference in the denominator at the same time is getting smaller providing the
 very  small IR enhancement we noticed for the  product of $\sigma_{s}$ with the MT model. Since for the other amplitude 
the mass function is replaced by one ($M \rightarrow 1$) which is larger than $m_{q}+M_{\Sigma} $, and since the denominator
 is the same, we expect for the same reasons as before to have  stronger  than the $\sigma_{s}$ IR support for the MT model. 
For the heavy quarks we  assume  $M_{H}\sim 2 m_{q}$.  Then 
\begin{eqnarray}
\sigma_{s}(q^{2}\sim 0)\sim \frac{1}{A} \cdot \frac{M_{\Sigma}+m_{q}}{M_{\Sigma}^{2}+2m_{q}M_{\Sigma})}
\label{Eq:sigma_s_Real_q_approx_IR_heavy}
\end{eqnarray}
  We keep the $M_{\Sigma}^{2}$ in the 
denominator because now it is of the order of 1 GeV.  A has values close to one and does not affect much the behavior
 of the amplitude near the peak. Because we  now have the sum of two positive terms in the denominator, we expect that 
will lightly decrease  the values of $\sigma_{s}$ near the peak. The large quark mass  in the numerator inhibits stronger
 suppressing IR effects from this amplitude but something similar does not exist in  $\sigma_{v}$ resulting in a  stronger 
 IR suppression. These observations help us to qualitatively understand the different IR effects we noticed in the last 
figures and how and why they change as the quark mass increase. We should notice at this point that the behavior of 
 $\sigma_{s}$ in the IR region is model dependent and for the \textit{heavy quarks} may not present a realistic
 behavior, but  for the other amplitude, $\sigma_{v}$, we know for certain there will be a 1/$m_{q}$ suppression for
 the \textit{heavy quarks} as we increase the quark mass.
  
The UV region analysis is easier since we are in the perturbative region and for both light and heavy quarks we have 
$m_{q}+M_{\Sigma}\sim m_{q}$ and $A\sim 1$. This is essentially model independent analysis.  Notice at this point that 
for  our analysis we are talking about dressing effects in the quark mass \textit{in} the IR or UV region
 \textit{originating} either from IR or UV interaction effects. There is though a shift in the momenta for the
 propagator amplitudes so in reality, when we refer to  mass dressing \textit{in the IR region}, we actually
 refer to the dressing \textit{in} the quark mass \textit{near the peak} (from $Re(q^{2}_{\pm})=-(\eta M)^{2}$ 
to about $Re(q^{2}_{\pm})\sim-(\eta M)^{2}+2$  $GeV^{2}$ on the real axis) of the meson mass shell parabolic region in
 the complex plane \textit{due to IR and/or UV interaction effects}, and  when we talk about mass 
dressing \textit{in the UV  region}, we actually mean  the dressing \textit{in} the quark mass 
\textit{far from the peak} (approximately for  $Re(q^{2}_{\pm})>-(\eta M)^{2}+2$  $GeV^{2}$  on the real axis)
  of the meson mass shell parabolic region in the complex plane
 \textit{due again to IR and/or UV interaction effects}. 
The idea of a quark receiving, at different momentum scales, both IR and UV dressing \textit{at the same time},
  can be accommodated by  the wave-like nature of the particle. There is  an insignificant quark mass dressing \textit{in}
 the UV region (for $q^{2}>4-5~GeV^{2}$ which for the propagators momenta in the BSE is for 
 $Re(q^{2}_{\pm})>-(\eta M)^{2}+5$  $GeV^{2}$  on the real axis in the corresponding parabolic region) for \textit{all} 
quark masses and that's why we can assume the last approximation for the quark mass function. The reason for this is that
  the internal quark propagator momentum in the gap equations depend on the external momentum $p^{2}$ and as the last  
one increase provides through the propagator in the gap equations integrand an increasing suppression on IR and UV 
dressing effects for   the quark mass \textit{in} the UV region. For the same reason the MT model provides less 
quark binding as  $p^{2}$ increase and we move to the UV region, the gluon momentum in the BSE is $k=p-\tilde{q}$ 
so we have an  $e^{-p^{2}/\omega^{2}}$ term strongly suppressing IR  binding forces \textit{in} the UV region and 
through a more complicated dependence of the UV tail on $p^{2}$ we also have suppression of the UV binding effects 
again \textit{in} the UV region.  The physical interpretation is that as the quark momentum $p^{2}$ increase the particle 
will be more localized, receiving less IR and UV dressing and at the same time feel less of  the color field of 
the other quark.  
We have then 
\begin{eqnarray}
\sigma_{s}(q^{2}> 5~GeV^{2})\sim \frac{m_{q}}{q^{2}+m_{q}^{2}-M_{H}^{2}/4} 
\label{Eq:sigma_s_Real_q_approx_UV}
\end{eqnarray}
For light quarks the numerator is very small and as $q^{2}$ increase the amplitude will decrease very fast. For 
$\sigma_{v}$ though the numerator is one which is larger than the quark mass and that decrease will be weaker.
For the heavy quarks on the other hand since  $M_{H}\sim 2 m_{q}$  the amplitude is further simplified to 
  $\sigma_{s}(q^{2}> 5~GeV^{2})\sim m_{q}/q^{2}$. As the current mass further increase will provide more support
 for the values of  $\sigma_{s}$ in large momenta. If we replace that mass with one to get the other amplitude, 
since now the quark masses are larger than 1 GeV, there will be less UV enhancement than  with the  $\sigma_{s}$. 
The simple form of the denominator also makes possible for the product of the two functions with the MT model to 
follow closer the changes in the behavior of the model, in fig. (\ref{fig:MT.sigma_s_2_new_ss_cc_bb_ms_0.083_mc_0.88_mb_3.8},
 \ref{fig:MT.sigma_v_2_new_ss_cc_bb_ms_0.083_mc_0.88_mb_3.8}) notice the obvious bending at about 2-3 $GeV^{2}$ of 
the c and b quark cases of products. 
Finally notice that we have very \textit{strong increase} in the suppression effects in the IR region as the quark
 mass is raised and that affects the propagator amplitude products in a \textit{small} area near the tip of the 
parabolic region, while there is a \textit{comparatively mild decrease} in the suppression effects in the UV
  region as the quark mass is raised in a \textit{comparatively much larger} area of the mass shell parabolic
 region in the complex plane. Therefore, based on our analysis at the end of previous subsection B,  
the \textit{integrated strength}  of the \textit{increase}  in the UV  dressing and binding
 effects should be  \textit{slightly more} that the  \textit{integrated strength} of the \textit{decrease} 
in the IR  dressing and binding effects  so that at the end  will have an  \textit{overall increase} in the quark
 mass UV dressing and binding energy \textit{in both} IR and UV regions.

 \subsection{Physical interpretation and other aspects of the relative effects of the UV and IR dynamics.}

Summing up, we found that as the current quark mass increase the propagator amplitude  $\sigma_{v}$ and less  $\sigma_{s}$
 will suppress mass dressing and binding IR contributions while initially for light quarks $\sigma_{v}$ will enhance IR 
contributions. At the same  time  $\sigma_{s}$ will provide the support that will enhance UV dressing and binding effects 
as the current mass is raised. From the location of the singularities of the c quark propagator, from the full model 
calculations and when we keep the IR or UV term (see Table \ref{Table:Full_IR_UV_approximate_singularities_location}), 
and from  plots 
(\ref{fig:MT.sigma_s_2_new_ss_cc_bb_ms_0.083_mc_0.88_mb_3.8},  \ref{fig:MT.sigma_v_2_new_ss_cc_bb_ms_0.083_mc_0.88_mb_3.8}) 
qualitatively we may conclude that this is the current quark mass region where this transition, in dominance of dressing and binding 
from UV over  IR region contributions, takes place. In terms of physics, the long distance interaction between the quarks (IR term) 
is inhibited by their  large mass and now they interact mostly through the exchange of short-range (large  momentum) gluons. 
Notice at this point the unique feature of the theory where the interaction degrees of freedom (gluons) depend on the mass 
of their source.
As the mass is raised  becomes increasingly more difficult for the quarks  to move further apart with  the large current
 quark mass gradually  replacing in that way the effect of the strong IR term of the interaction. In other words, the 
increase in the current mass  now replaces the strong quark mass dressing   IR effects of the model (enhancing at the same time UV 
dressing), making that term  almost unnecessary, and with only a small attraction we can have a bound state. 

From the c quark amplitude  plot in fig. (\ref{fig:Re_sigma_s_1st_MT_term_kcp_32_32_32_32_E-7_P2-8.0_n_0.5_0.88}) and a 
similar one for the b quark amplitude,  becomes obvious that the MT kernel, because of the behavior of the infrared term, 
 does not support a single pole-mass constituent-like  behavior and actually, from the b quark  plot, it appears  we are 
moving further away from that type of behavior as the current quark mass increase.  We can assume then that a  
constituent-like   behavior can be supported by the effective kernel, as we increase the quark mass, only if there is
 an \textit{important decrease} in the strength of the IR term of the kernel.

For the solution of the gap equation one starts by assuming a free quark boundary condition at some large space-like 
momentum scale, well inside the perturbative region, and this solution is modified because of the interaction of the quark 
with the vacuum, mostly in the nonperturbative region for the light quarks. If we ignore the strong nonperturbative effects,
 expressed through the first term of the MT kernel, and keep only the perturbative UV tail term,  
then from perturbative QCD we know that 
the  pole of the propagator,  will retain some  of its initial features determined by the current quark mass. 
The fact that we still have a first singularity on the real axis 
(fig. \ref{fig:Re_sigma_s_2st_MT_term_kcp_32_32_32_32_E-7_P2-8.0_n_0.5_0.88} and similar plot for the b quark) is a 
verification of that. On the other hand as we go to heavier quarks  the IR dressing effects will become  comparatively 
smaller, the location of the singularities  will be mostly determined by the current quark mass (and UV dressing) and they will be very 
close, along the real axis, to the mass-pole singularity of the free propagator. Therefore the  dynamics of the
 interaction,  as expressed through the MT model (specifically the first part) , will have a lesser  role in
 the \textit{location} of the  singularities as the current quark mass is raised, but will  be always 
responsible for their \textit{type} 
(confinement excludes poles) which still has a very subtle and important  role for the calculation of certain 
meson observables.   Therefore we expect  that  a small variation of the parameters of the model,
will have insignificant impact in the location of the heavy quark propagator singularities.
As a consequence the  values of the physical observables will also show indifference to these variations and will be 
mostly determined by the quark masses. Reversing the reasoning, if the IR term of the model had an explicit quark mass
 dependence, we should expect a decreasing dependence of the strength of the model on the  mass as we go to heavier quarks. 


\section{An effective kernel with quark mass dependence.}
Higher order diagrams in the expansion of the BSE kernel contain  internal quark propagators which will introduce a quark
 mass dependence to it. The  parameters of an effective kernel that is applicable over a wide range of quark mass can be 
expected to have an explicit quark mass dependence.   The way the parameters should vary with the quark mass is not easy 
to determine. In ref. \cite{Eichmann:2008ae} it was noted that  diagrams higher order than ladder-rainbow provide an
 attractive effect on mesons which decreases with increasing quark mass. Qualitative estimates of this effect for quarkonia 
were used to produce an $m_{q}$-dependent effective kernel (of rainbow-ladder format)  whose  role was to produce meson
 bound state quantities that left room for subsequent and explicit corrections from higher order kernel processes. The 
IR strength and range of such a  core  kernel was fitted to reproduce the expected behavior of  eq.(9) in  
ref. \cite{Eichmann:2008ae} in the low quark mass region. With  the MT-model form used as a template for the core  
kernel, the IR strength  was fitted to light meson properties leaving room for  higher order terms effects. On the
 other hand the IR strength of the MT effective kernel was fitted to light quark meson properties absorbing higher
 order effects. Since the MT-model parameters are fixed these effects will be still present in the heavy quarks region. 
The extrapolation of the core kernel to the heavy quark region provides an interesting point of comparison with the MT-model.

 So far we had numerous indications that the MT effective kernel provides too much dressing for the heavy quarks. 
At the same time we believe it gives a BSE kernel that is too attractive and that cancels to some degree the 
overdressing effects in the quark mass. To make things more complicated, we found that the heavy quark propagator 
itself weakens to a certain degree the model's IR effects. Therefore it is not so easy to determine the net effect of the  MT 
interaction kernel in the heavy quark region from the calculated meson observables.  Since  the 
heavy quark region was not considered in ref. \cite{Eichmann:2008ae}, we extend its application in the c and b quark region
and we examine any benefits for studying heavy quark mesons.

The expression for the core effective kernel of  ref. \cite{Eichmann:2008ae} is:
\begin{eqnarray}
\frac{4 \pi \alpha(k^{2})}{k^{2}} = C(\omega,\hat{m})\frac{(2\pi)^{2} k^{2}} {\omega^{7}} e^{-\frac{k^{2}}{\omega^{2}}}+ 
 \nonumber  \\
\frac{2 (2\pi)^{2}  \gamma_{m} F(k^{2})}{\ln[\tau+(1+\frac{k^{2}}{\Lambda^{2}_{QCD}})^{2}]} .  
\label{Eq:MT_core_kernel}
\end{eqnarray}
 The midpoint value of  $\omega$ in the minimal sensitivity region,  from  studies going as far as about the $s$ quark 
mass, was found to have the following dependence on the renormalization-group-invariant quark current mass $\hat{m}$:
\begin{eqnarray}
\omega(\hat{x})=0.38+\frac{0.17}{1+\hat{x}},~~\hat{x}=\frac{\hat{m}}{\hat{m}_{0}},~~\hat{m}_{0}=0.12~GeV ,
\label{omega_quark_mass_dependence}
\end{eqnarray}
and in this case the product of the two core-model parameters, D and $\omega$,   will  depend only on $\hat{x}$:
\begin{eqnarray}
C(\hat{x})=\omega D= C_{0}+\frac{0.86}{1+C_{2}\hat{x}+C^{2}_{3}\hat{x}^{2}}
\label{C_quark_mass_dependence}
\end{eqnarray}
with $C_{0}=0.11$, $C_{2}=0.885$, $C_{3}=0.474$.

\begin{figure} 
\centering
\includegraphics[height=5cm,width=8cm]{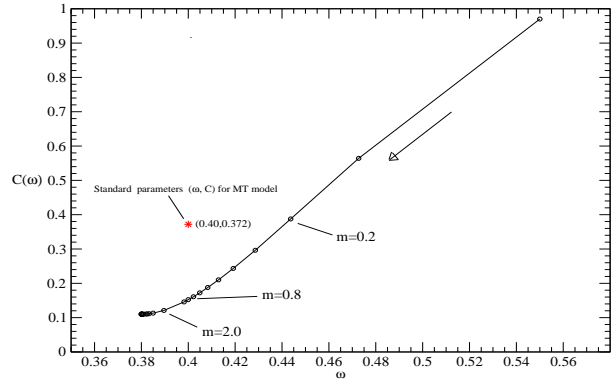}
\caption{Variation of  parameter $C(\omega)=\omega D(\omega)$  vs. parameter $\omega$ of the core kernel of ref. 
\cite{Eichmann:2008ae} for $0\leq \hat{m}<+\infty$. The arrow points in the direction of increasing  $\hat{m}$.}
\label{fig:C_vs_OMEGA_MT_parameters1}
\end{figure}

If we assume these  relations are true for all current masses then we notice that from the chiral limit all the way to
 infinite quark mass, parameter  $\omega(\hat{x})$ varies from  $\omega(0)=0.55~GeV$ to  $\omega(+\infty)=0.38~GeV$ and
 $C(\hat{x})$ from  $C(0)=0.97~GeV$ to  $C(+\infty)=0.11~GeV$ or for $D(\hat{x})$ from  $D(0)=1.763$ to  $D(+\infty)=0.289$. 
So  the infinite range of quark masses is mapped into a very small domain of the parameters   $\omega(\hat{x})$ and  
$C(\hat{x})$. From the plot in fig. (\ref{fig:C_vs_OMEGA_MT_parameters1}) we observe there is a very fast variation of 
$C(\omega)$ for quark  masses up to $\hat{m}\sim 2-3~GeV$ and then above that  the parameters rapidly approach  their 
limit values and they don't vary much.  
The limiting kernel parameters C, $\omega$ are a possible definition of a rainbow -ladder kernel for very massive quarks.
 As a qualitative comparison with the MT-model at high quark mass we present some meson results from use of the new  
effective kernel which we call a core model.

Since the $c$ quark mass of 0.88 GeV at scale $\mu=19~GeV$ corresponds to a $\hat{m}$ of about 1 GeV, we obtain  
$D=0.308$ and $\bar{\omega}=0.39~GeV$   \cite{comm1}   for this core kernel. For the $b$ meson  studies we use the 
core model limit  values (D=0.289, $\omega=0.38~GeV$).


\begin{table*}
\caption{$c$ and $b$  quarkonia masses and decay constants using the values  from the MT and the  core model \cite{Eichmann:2008ae}. 
The letter S  signifies the MT-kernel  values of parameters D, $\omega$. For reasons of comparison with the core model results in the last row for the b quarkonia 
we also include the results from the UV tail term only.
Experimental data are also in GeV. \label{Table:cc_bb_D_w_kcp_PS_V_meson_N_f}} 
\begin{center}
\begin{tabular}{|l||l|l||l|l|} \hline \hline   
c Quarkonia &\multicolumn{2}{c||}{$(M^{exp.}_{\eta_c},f^{exp.}_{\eta_c})=(2.980,0.340)$} & \multicolumn{2}{c|}{$(M^{exp.}_{J/\psi},f^{exp.}_{J/\psi})=(3.097,0.416)$ } \\  \hline 
 (D, $\omega$(GeV))  & $M_{\eta_c}$(GeV) &$f_{\eta_c}$(GeV) &$M_{J/\psi}$(GeV)& $f_{J/\psi}$(GeV)    \\  \hline  \hline     

 (0.93,0.40)S   & 3.035       & 0.387     & 3.235    & 0.415      \\  \hline 
 (0.308,0.39)  & 2.934       & 0.334     & 3.050    & 0.319      \\  \hline \hline \hline 

b Quarkonia &\multicolumn{2}{c||}{$(M^{exp.}_{\eta_b},f^{exp.}_{\eta_b})=(9.30,-)$} & \multicolumn{2}{c|}{$(M^{exp.}_{\Upsilon},f^{exp.}_{\Upsilon})=(9.46,0.700)$ } \\  \hline  \hline 
 (D, $\omega$(GeV))  & $M_{\eta_b}$(GeV)&$f_{\eta_b}$(GeV) &$M_{\Upsilon}$(GeV)&$f_{\Upsilon}$(GeV)   \\  \hline  \hline    
 (0.93,0.40)S   & 9.585       & 0.692     & 9.685    & 0.682     \\  \hline 
 (0.289,0.38)   & 9.537       & 0.633     & 9.612    & 0.589     \\  \hline  
 UV only        & 9.530       & 0.606     & 9.586    & 0.510     \\  \hline \hline

\end{tabular}\\[2pt]
\end{center}  
\end{table*}
The results for the c- and b-quarkonia masses and their decay constants from the core model for  comparison with the MT 
model results   are collected in table (\ref{Table:cc_bb_D_w_kcp_PS_V_meson_N_f}).
Summing up, the relative percentage changes in the parameters of the model and the resulting relative percentage changes
 in the quarkonia masses and decay constants,  for the  $c$ quarkonia  are:  
\begin{eqnarray}
 (\Delta D/D,\Delta \omega/\omega)=(-66.88,-2.5)\%   \nonumber \\
 \             \rightarrow (\Delta M/M,\Delta f/f)_{\eta_c}=(-3.3,-13.7)\%   \nonumber \\
               \rightarrow (\Delta M/M,\Delta f/f)_{J/\psi}=(-5.7,-23.1)\%  \nonumber                                    
                                                                 \label{cc_D_w_M_f_relative_prcg_dif}
\end{eqnarray}
 while  for the $b$ systems:

\begin{eqnarray}
(\Delta D/D,\Delta \omega/\omega)=(-68.92,-5.0)\%    \nonumber \\
                       \rightarrow (\Delta M/M,\Delta f/f)_{\eta_b}=(-0.50,-8.5)\%   \nonumber \\  
                       \rightarrow (\Delta M/M,\Delta f/f)_{\Upsilon}=(-0.75,-13.6)\% .    \nonumber                       
                                                                              \label{bb_D_w_M_f_relative_prcg_dif}
\end{eqnarray}
It is obvious that the decay constants, especially those of the vector mesons, since they are more extended particles, 
are much  more sensitive to parameters changes. Vector meson masses also appear to be relatively more sensitive 
than the pseudoscalar ones, for  $b$ quarkonia though  the  relative mass change 
 indicate  insensitivity to the IR  dynamics of the model. Finally by comparing the core model results for the b quarkonium with
those  obtained with  the UV term only in the model, we see that  masses, especially that of the vector meson, are closer to the experimental  ones but the  vector meson decay constant at the same time becomes even  smaller.


\section{Recovery of a \lowercase{q}Q meson mass shell.}

A qQ meson mass shell can not be obtained with a  full model dressing for both quarks, and that reveals the limitations of
the approach for light-heavy meson studies.  
Given that the core model of an effective BSE kernel is $m_{q}$-dependent, and that its use for qQ mesons containing  
quarks of different masses is not defined, we decided to explore the point of view that the DSE kernel for the dressing 
of the heavy quark propagator could be too strong in the infrared.  Hence we performed qQ meson calculations in which the 
IR term of the MT-model kernel is removed. 
The BSE kernel for the interaction of the light and heavy quark though should be approximately unchanged since there is no
 significant change in the size of the qQ meson compared to the size of the light quark  mesons. Solving the DSE  for the
 b quark propagator with only the UV tail term  of the MT kernel, where   the full MT kernel was used  for the calculation
  of the light quark propagator and  the solution of the BSE, we were able to reach a mass shell  for light-heavy mesons
 having a  b quark. The masses and decay constants appear in Table 
(\ref{Table:qb_q_BSE_MT_b_UV_PS_kcp_new_meth_meson_masses_decays}). These calculations represent a suppression 
of IR dressing of heavy quarks in mesons. 

\begin{table}
\caption{qb q=u/d,s,c   pseudoscalar meson masses and decay constants with their  differences from experiment after
 eliminating  the IR term of the MT kernel    in the  dressing of the b-quark. The full MT kernel was used for the
 light quark propagator and the solution of the BSE. All data are in GeV units.
\label{Table:qb_q_BSE_MT_b_UV_PS_kcp_new_meth_meson_masses_decays} }
\begin{center}
\begin{tabular}{|l||l||l|l|l||l|l|l|} \hline \hline
qb   & $\eta$ & $M^{exp.}_{H}$ & $M^{UV}_{H}$ &$\Delta M/M\%$ & $f^{exp.}_{H}$&$f^{UV}_{H}$&$\Delta f/f\%$ \\ \hline   
   
ub  & 0.90      & 5.279        & 4.658         &$-12.16$       & 0.176                  &0.133       &$-24.4$  \\  \hline 
sb  & 0.90      & 5.370        & 4.748         &$-11.59$       &--                      &0.164       &--       \\  \hline 
cb  & 0.86      & 6.286        & 5.831         &$-7.238$       &--                      &0.453       &--       \\  \hline  \hline  

\end{tabular}
\end{center}
\end{table}
 
With this modification of the dressing of the b-quark  we can  find meson  masses that are only about 7-12 \% smaller than 
the experimental ones. For  the only decay constant experimentally known there is a -24.4 \% difference. 
The present results and the  earlier results  for the equal quark systems appear to provide a partial confirmation of the
 recent suggestion  by  Brodsky and Shrock of a   maximum wavelength for quarks and gluons in  mesons 
\cite{Brodsky:2008be},  \cite{Brodsky:2008xu}, \cite{Brodsky:2008xm}.
The existence of such a  wavelength for quark and gluons would be due to confinement in mesons   with size of the
 order of 1 fm. That in turn, through  the Compton relation, will introduce a minimum   quark and gluon momentum 
inside hadrons. Same 
reasoning can be applied to Baryons. These scales though will depend on the bound state of quarks, requiring different scales 
to be artificially introduced for each case and mathematically accommodate the Dyson-Schwinger equations solution.  If 
QCD is the correct theory for the interaction of quarks and gluons something like that is not desirable.           
 The present studies   may  indicate a possible scenario for the natural realization of these scales  through the combined 
IR dynamics of the   quark mass dependent  kernel and the quark propagators  in full QCD calculations. 
Some simple preliminary studies qualitatively support  that idea but further and more detailed investigation is required.

\section{Conclusions.}  

The size of the QQ mesons becomes smaller as quark mass increases and the dressed quark quasi-particles themselves become  
smaller in size. The infrared component of the kernel relates to large distance gluons; such components should be less 
physically relevant for internal dynamics of heavy, small Compton size, particles. This is supported by our finding that 
the heavier QQ quarkonia receive diminishing  contributions from the infrared component of our model kernel; the UV 
component alone provides a very good description of the $b\bar{b}$ states. We stress  that this refers not only to the
 binding interaction but also to the quark self-energy dressing. 

In detail, as the quark mass increases, due to its  smaller Compton size, UV  dressing will  increase 
and IR dressing will 
be strongly suppressed.  Model independent evidence  indicate  that the short distance dressing will be, after certain mass
scale, more than the long distance dressing, and at a higher mass scale,  it may be over and above the long distance
dressing of the lightest u/d   quarks.
   The heavy quarks will be limited in a
smaller area in space and the binding force  will be provided by the exchange of sort distance gluons,
 signifying a smaller overall binding energy.  The heavy quark mass and the now relatively stronger UV dressing,
\textit{triggered by} that large mass, make feasible, with a  very weak force, to bind the particles together.

On the other hand, the qQ mesons have a size that does not diminish significantly with increasing heavy quark mass. So the 
infrared sector of the binding interaction  should remain relevant. However the self-energy dressing of the heavy quark 
should not receive strong contributions from large distance gluons. This becomes evident  by our finding  that a 
suppression of the infrared component of the b-quark dressing kernel allows a physical $qb$,  $q=u/d,s,c$  meson state 
even though the binding effective interaction and the dressing of the light quarks remains unchanged. 

In a similar way  as the quark momentum increases the particle will become  more localized, receiving
 \textit{simultaneously} smaller  infrared \textit{and}  ultraviolet dressing  and feeling less 
of the color field of the other quark(s) in the  bound state. The wave-like nature of quarks  can accommodate that 
interpretation.

Aspects of this study may be  related to  Brodsky and Shrock's suggestion of a  maximum wavelength of quarks and gluons in
 hadrons. The present work   may actually  suggest a possible scenario for the natural realization of 
 such scales in QCD. Further studies  are necessary in that direction.


\begin{thebibliography}{18}
\bibitem{comm1}
Although we use an approximate relation to calculate $\hat{m}$ and there is some uncertainty,  the parameters vary slowly in that mass area and we don't have to worry  much about the exact value of the quark mass corresponding to these parameter values.
\end{thebibliography}
\end{document}